\documentclass[useAMS,usenatbib]{mn2e}

\usepackage{graphicx}
\newcommand{\dm}{$\Delta m_{15}$(B)}

\title[Spectral diversity of Type Ia Supernovae]{Spectral diversity of Type Ia Supernovae}
\author[James, Davis, Schmidt and Kim]{J. Berian James$^{1,2}$\thanks{E-mail:
jbjames@physics.usyd.edu.au (JBJ); tamarad@mso.anu.edu.au (TMD); brian@mso.anu.edu.au (BPS); 
AGKim@lbl.gov (AGK)}, Tamara M. Davis$^{1}$, Brian P. Schmidt$^{1}$
 and Alex G. Kim$^{3}$\\
$^{1}$Research School of Astronomy and Astrophysics, Australian National University, ACT 0200 Australia\\
$^{2}$School of Physics, University of Sydney, NSW 2006 Australia\\
$^{3}$Physics Division, Lawrence Berkeley National Laboratory, 1 Cyclotron Road
Berkeley, CA 94720 USA}
\begin{document}

\date{}

\pagerange{\pageref{firstpage}--\pageref{lastpage}} \pubyear{2006}

\maketitle

\label{firstpage}

\begin{abstract}
We use published spectroscopic and photometric data for 8 Type Ia supernovae to construct a dispersion spectrum for this class of object, showing their diversity over the wavelength range 3700~\AA\ to 7100~\AA. We find that the B and V bands are the spectral regions with the least dispersion, while the U band below 4100\AA\ is more diverse.  Some spectral features such as the Si line at 6150\AA\ are also highly diverse. We then construct two objective measures of `peculiarity' by (i) using the deviation of individual objects from the average SN Ia spectrum compared to the typical dispersion and (ii) applying principle component analysis.  We demonstrate these methods on several SNe Ia that have previously been classified as peculiar.
\end{abstract}

\begin{keywords}
supernovae : general
\end{keywords}

\section{Introduction}\label{intro}

Type Ia supernovae (SNe Ia) are observed to display remarkably homogeneous light curves in broadband photometry.  They are currently classified according to a single-parameter family of peak luminosity as a function of light curve shape~\citep{phil93}, which allows them to be used as accurate cosmological distance indicators. 

However, we know that this one-parameter family does not completely encompass the variety of SNe Ia observed. Once we have reduced any contribution from measurement error to the point where its effects are negligible the intrinsic diversity of SNe Ia themselves represent the limiting accuracy to which a one-parameter family can aspire.

This diversity manifests itself as differences in the SN Ia spectra.
It is important to assess the diversity in particular features of SN Ia spectra and find those hot spots that have the most significant effect on the broadband photometry used for cosmology.  Identifying these points of variability in SN Ia spectra has a multi-fold benefit.  (1) It allows us to identify which spectral regions are the most stable for photometry and thus design future dedicated SN cosmology probes to utilise these stable wavelength regions.  (2) It allows us to classify SN Ia based on certain spectral features into sub-types that have a tighter magnitude-redshift relation than the entire SN Ia population.  
(3) It allows us to put weighted error bars on the points in the magnitude-redshift relation depending on which part of a redshifted SN Ia spectrum was sampled by the filter set in use. 

The quest to understand the diversity in the type Ia supernova population is accelerating, with many groups investigating this in an attempt to control systematic uncertainties associated with using type Ia supernovae as standard candles.  
Empirical methods for studying the variety in the type Ia population focus on the linked aspects of photometric diversity \citep{jha05} and spectroscopic diversity. 
Many of the spectral studies concentrate on fitting spectral lines (such as the Si II feature), measuring how they evolve with time, relating these to physical parameters such as metallically or ejecta velocity, and studying correlations between these features and the peak magnitude or light curve width \citep{nugent95,filippenko97,benetti04,blondin04,gallagher05,ben05}.  Efforts to compare these observational parameters to the results of theoretical models are ongoing \cite[e.g.][]{travaglio05,branch05,branch06,bongard06}, and this important topic promises to allow the prediction of potential changes in type Ia supernova demographics over time.  In the meantime observational correlations, such as the width-luminosity relation, remain the strongest methods for controlling the effects of supernova diversity on cosmological parameter estimation. 

In this paper we design an empirical measure of diversity that uses a new approach.  Instead of fitting spectral lines we subject the entire spectrum to a number of different statistical analyses.  These show promise for providing an objective method of distinguishing peculiar supernovae (at the very least) and for finding correlations amongst subsets of the type Ia population. In Section~\ref{sect:data_analysis} we describe the data set and our analysis technique.  In Section~\ref{sect:discussion} we provide plots of the spectral variability of SNe Ia as a function of wavelength and discuss the features.  In Section~\ref{sect:peculiarity} we provide a prescription for determining the degree of `peculiarity' of a supernova based on its spectrum near maximum light.

\section[]{Data Preparation}\label{sect:data_analysis}

The sample consists of 18 previously published spectra taken for 8 low redshift type Ia SNe. Table~\ref{table:spectra} lists the spectra and the source catalogues.  These represent the subset of published objects that are confirmed to be normal SNe Ia and have good quality spectra within two days of maximum light as well as established $\Delta m_{15}$ values, B magnitudes at maximum light ($M_{\rm B}$), extinction values ($E(B-V)$) and  distance moduli ($\mu_0$). Table~\ref{table:objects} lists the photometric properties of the source objects.

\begin{table}
\caption{Catalogue of SNe Ia spectra used in this study.\label{table:spectra}}
\begin{tabular}{crrrc}
\hline
SN&epoch (days)~&{$\lambda_{\rm min}$ (\AA)~~}&{$\lambda_{\rm max}$ (\AA)~~}&Reference$^a$~\\
\hline
 1981B & 0 & 1272 & 8392& Br83\\
 1989B & $-1$ & 3001 & 11167 & We94\\
 1992A & $-1$ & 3564 & 7100 & Ki93 \\
 1994D & $-2$ & 3402 & 9132 & Pa96\\
       &  $0$ & 3069 & 10130 & Br\\
       &  $2$ & 3466 & 9204 & Pa96\\
       &  $2$ & 4439 & 7013 & Pa96\\
 1994S &  $0$ & 3120 & 11300 & Br\\
 1996X & $-2$ & 3748 & 9917 & Sal01 \\
       &  $0$ & 3079 & 10670 & Sal01 \\
       &  $1$ & 3078 & 10669& Sal01 \\
 1998aq&  $0$ & $3720$ & $7500$ & Br03\\
       &  $1$ & 3720 & 7511  & Br03 \\
       &  $2$ & 3720 & 7521 & Br03 \\
1999ee & $-2$ & 3262 & 9983 & Ha02 \\
       & $-1$ & 3559 & 9134 & Ho \\
       &  $0$ & 3262 & 9687 & Ha02 \\
\hline
\end{tabular}
\medskip\\
$^a${\sc References.}---Br83: Branch et al.\ 1983; Br03: Branch et al.\ 2003; Ha02: Hamuy et al.\ 2002; Ki93: Kirshner et al.\ 1993; Pa96: Patat et al.\ 1996; Sal01: Salvo
\end{table}

\begin{table*}
\begin{minipage}{126mm}
\caption{Catalogue of photometric parameters for SNe Ia.}\label{table:objects}\begin{tabular}{crrrrrrl}
\hline
SN&{$\Delta m_{15}(B)$}&$B_{\rm max}$~~~&{$V_{\rm max}$~~~}&{$E(B-V)$}&{$\mu_0$~~~~~~}&{References$^b$}\\
\hline
 1981B & 1.10(07) & 12.00(02) & $-~~~~~$ & 0.11(03) & 30.10(30) & Br83\\
 1989B & 1.31(07) & 12.34(05) & 12.02(05) & 0.34(04) & 29.73(04) & We94 \\ 
 1992A & 1.47(05) & 12.60(20) & 12.61(05) & 0.00(02)& 31.59(05) & Ph99; Le93 \\
 1994D & 1.32(05)& 11.84(05) & 11.92(05) & 0.00(02) & 31.10(07) & Pa96; Ph99 \\
 1994S & 1.10(10) & 14.85(04) & 14.84(06) & 0.00(03) & 34.36(03) & Kn03; Ph99 \\
 1996X & 1.31(08) & 13.24(02) & 13.21(01) & 0.01(02) & 32.19(07) & Ph99; Sal01 \\
 1998aq & 1.14(10) & 12.39(05) & 12.57(05) & 0.014(05) & 31.72(14)& Sah01; Br03 \\
 1999ee & 0.94(06) & 14.93(02) & 14.68(15) & 0.30(04) & 33.55(23) & Ha02; St02 \\
\hline
\end{tabular}
\medskip\\
$^b${\sc References.}---Br83: Branch et al.\ 1983; Br03: Branch et al.\ 2003; Ha02: Hamuy et al.\ 2002; Kn03: Knop  et al.\ 2003; Le93: Leibundgut et al.\ 1993; Pa96: Patat et al.\ 1996; Ph99: Phillips et al.\ 1999; Sah01: Saha et al.\ 2001; Sal01: Salvo et al.\ 2001; St02: Stritzinger et al.\ 2002; We94: Wells et al. \ 1994
\end{minipage}
\end{table*}

Correction for dust extinction is made according to the standard extinction law of\citet{savage79}, using colour excess values $E(B-V)$ from \citet{phil99} where available.

It is a familiar task to convert apparent magnitudes to absolute magnitudes in order to compare the brightness of objects at different distances.  In order to compare the 18 spectra in our sample we have to perform the spectral equivalent of converting from apparent to absolute magnitude. We do this by multiplicatively scaling the spectra so they give the correct absolute magnitude in the B or V band after synthetic photometry.   

There is more than one way to determine the `correct' absolute magnitude.  We could choose to scale each spectrum to the broadband magnitude of its source supernova, once corrected for extinction, distance modulus and \dm.  However the observational uncertainty on nearby SN Ia magnitudes is typically about 0.2 magnitudes, while the standard deviation in magnitude values is typically about 0.4 and 0.2 magnitudes in B and V respectively \citep{hamuy96}. This makes it difficult with the current data to distinguish broadband magnitude variations due to intrinsic spectral variability from variations due to observational error.

Instead we choose to scale each spectrum to a fixed broadband magnitude.  This allows us to probe the spectral diversity without contamination from observational magnitude uncertainties.  When we scale to a fixed magnitude we artificially minimise the diversity in that region. To examine the magnitude of this effect, we scale the spectra to fixed $M_B$ in the first instance and to fixed $M_V$ in the second. The choice of scaling has a small effect compared to the size of the spectral variations.  The mean values of the B and V absolute magnitudes are $M_{\rm B}=-19.16\pm0.05$ and $M_{\rm V}=-19.11\pm0.02$ ($1\sigma$), agreeing with the better performance of the \citet{hamuy96} relation in the V band.  

Scaling in a single band is achieved by allowing for a
linear relationship between the synthetic photometric magnitude and the
logarithm of the scaling factor (luminosity)\footnote{A refinement of this technique is to fit to both $M_B$ and $M_V$
simultaneously, which creates a wider, more shallow dependence of
variability on wavelength band, further reducing the intensity
relative to the spectral variations. This is achieved by minimising
a best-fit function for fixed magnitudes $B_c$ and $V_c$,
\begin{eqnarray}
\epsilon(k) & = & \left(B - B_{c}\right)^2 + \left(V - V_{c}\right)^2 \\
& = & (a\log k + b - B_{c})^2 + \left(c\log k + d - V_{c}\right)^2.
\end{eqnarray}
Differentiating with respect to $\log k$ and setting the
result equal to zero yields
\begin{equation}
k = \exp\left\{\frac{a(B_{c} - b) + c(V_{c} - d)}{a^2 +
c^2}\right\}\label{eq:predict}.
\end{equation}
The values of the coefficients are found as in the single-band case.
Equation~\ref{eq:predict} allows us to determine \emph{a priori} the
scaling factor required to fix a spectrum near a chosen magnitude in both
$B$ and $V$.},
\begin{equation}
M = a\log k + b\label{eq:mk},
\end{equation}
The value of the slope constant $a$ is constant for all
objects within a particular band. The offset $b$ varies depending on the units
of the spectrum, and is determined for each object separately by
computing the relation~\ref{eq:mk} for a range of scaling factors $k$. From this
follows the relationship for $k$ in terms of a fixed desired magnitude $M$.

This process provides a set of extinction-corrected SN Ia spectra that represents the most homogeneous spectral set we are able to generate.  We now test these spectra for diversity in the sample. 

\section[]{Measuring SN Ia Diversity}\label{sect:discussion}
For each object we compile a single spectrum based on the average of all available spectra for that object between $-2$ and 2 days from maximum. We use logarithmic bin spacing to offset the Doppler broadening of linewidths at higher wavelengths. The eight average spectra are binned to constant resolution, $R=\lambda/\Delta\lambda$, with bin size proportional to the bin centre,
\begin{equation}
\frac{\Delta\lambda}{\lambda}\approx\frac{v}{c} = \beta.
\end{equation}
Our choice of $\beta = 0.005$ corresponds to approximately 20\AA\ bins near 4000\AA\ and 35\AA\ bins near 7000\AA.
This choice of binning parameter is narrow enough to appropriately sample all prominent spectral features, but wide enough to be insensitive to noise, allowing for effective synthesis of the spectra.
We tested a variety of bin widths from $\beta=0.001$ to $\beta=0.02$ and the results are not sensitive to the width chosen.

\begin{figure*}
\begin{center}
\includegraphics[]{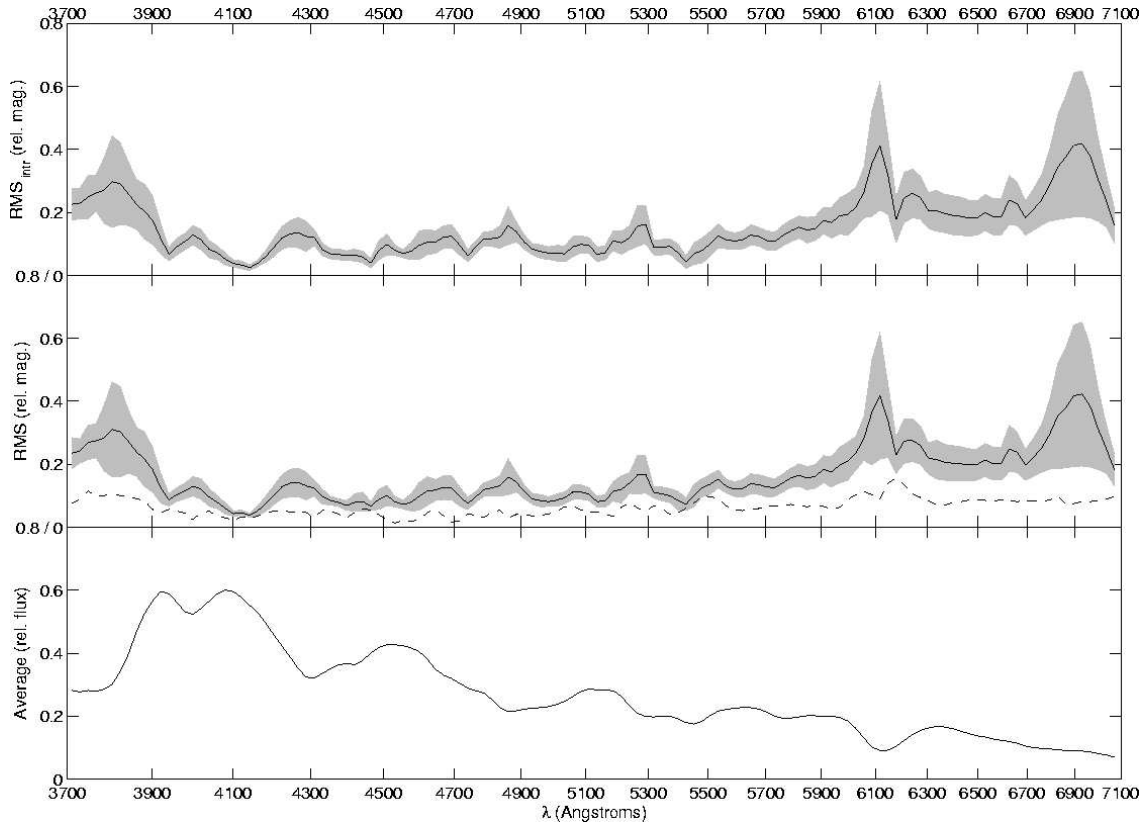}
\caption{To calculate the diversity in SN Ia spectra as a function of wavelength we first calculate the average spectrum (bottom panel) and the RMS deviation from this average (middle panel, solid line).  The uncertainty in this RMS is approximated by bootstrap resampling and is shown by the grey shading.  The dashed line in the middle panel is the intra-object dispersion, and represents our estimate of the contribution from observational uncertainty.  The upper panel shows our final result, which is the RMS minus this estimated observational dispersion.
In this plot the spectra have been scaled to a fixed magnitude in the B band.  This artificially lowers the scatter in the B-band region, centred around 4200\AA.  Large-scale shape variations in the spectra, would show up as a slow increase in the scatter away from this region (as we see here).  Errors in extinction correction would also have this effect.  
 \label{fig:scatter:fixedB}}
\end{center}\end{figure*}

\begin{figure*}
\begin{center}
\includegraphics[]{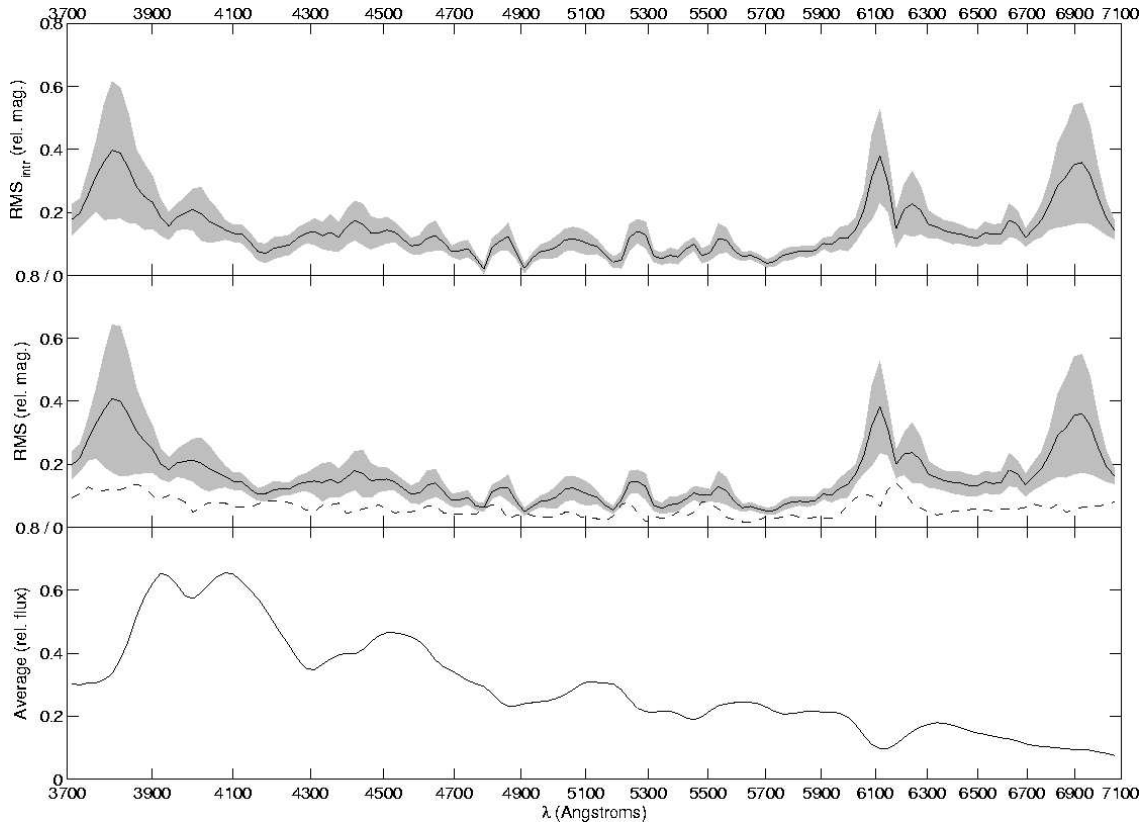}
\caption{As for Figure~\ref{fig:scatter:fixedB}, though here the spectra are scaled to have a the same magnitude in the V band (centred around 5200\AA), rather than the B band, as can be seen from the wavelength range of the region of lowest diversity. The same small-scale spectral features remain prominent, demonstrating the independence of the variance of these regions on the choice of anchor filter.  For example, the variable strength (and position) of the Si II line observed at around 6150\AA.
} \label{fig:scatter:fixedV}
\end{center}\end{figure*}

Using the averaged spectra from the eight supernovae in our sample we calculate the overall mean spectrum and determine the root-mean-square (RMS) deviation from the mean in each bin to quantify the diversity in that region.  The result is plotted in Figures~\ref{fig:scatter:fixedB} and~\ref{fig:scatter:fixedV}, where the lower panel shows the overall average spectrum and the solid line in the middle panel shows the RMS dispersion about the mean.  The shaded regions in this figure represent the uncertainty in the RMS dispersion, which is estimated by bootstrap resampling~\citep{efron82}.  

We wish to isolate the intrinsic diversity of SNe Ia from observational scatter. As the spectra used in the sample range from epoch -2 to 2 days we expect some diversity within the spectra for a single object. Moreover, differences between individual observations introduce scatter that is independent of intrinsic diversity in the SN Ia population. We seek to quantify this variance and remove it from the total, leaving only the intrinsic RMS.  We have already made a single spectrum for each object by taking the average of all its available spectra.  For each object, we account for the spurious dispersion by calculating the RMS variance of the available spectra for each object about the mean of that object. This quantity --- the intra-object variation --- when averaged over all the objects, gives an estimate of the dispersion that is not due to intrinsic differences between the supernovae. This quantity is also used as an estimate of the non-intrinsic dispersion for an object when only one spectrum is available.  This dispersion is the dashed line in the middle panel of Figures~\ref{fig:scatter:fixedB} and~\ref{fig:scatter:fixedV}. It is small compared to the dispersion between objects. Since the contribution from observational error is expected to only increase the scatter in the results we subtract (in quadrature) the observational scatter component from the overall RMS deviation.  The remaining RMS, ideally all due to intrinsic diversity, is shown in the upper panel of Figures~\ref{fig:scatter:fixedB} and~\ref{fig:scatter:fixedV}. 

\begin{figure*}
\begin{center}
\includegraphics[]{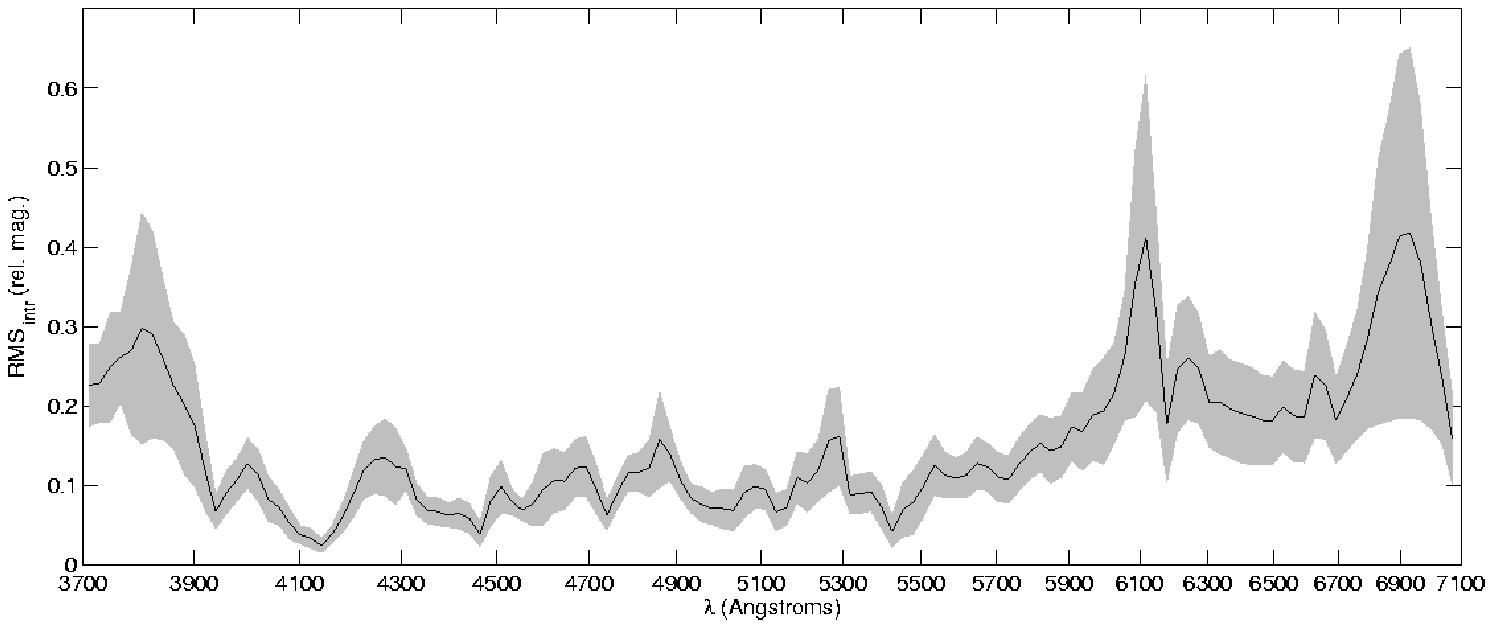}
\vspace{0.5cm}
\includegraphics[]{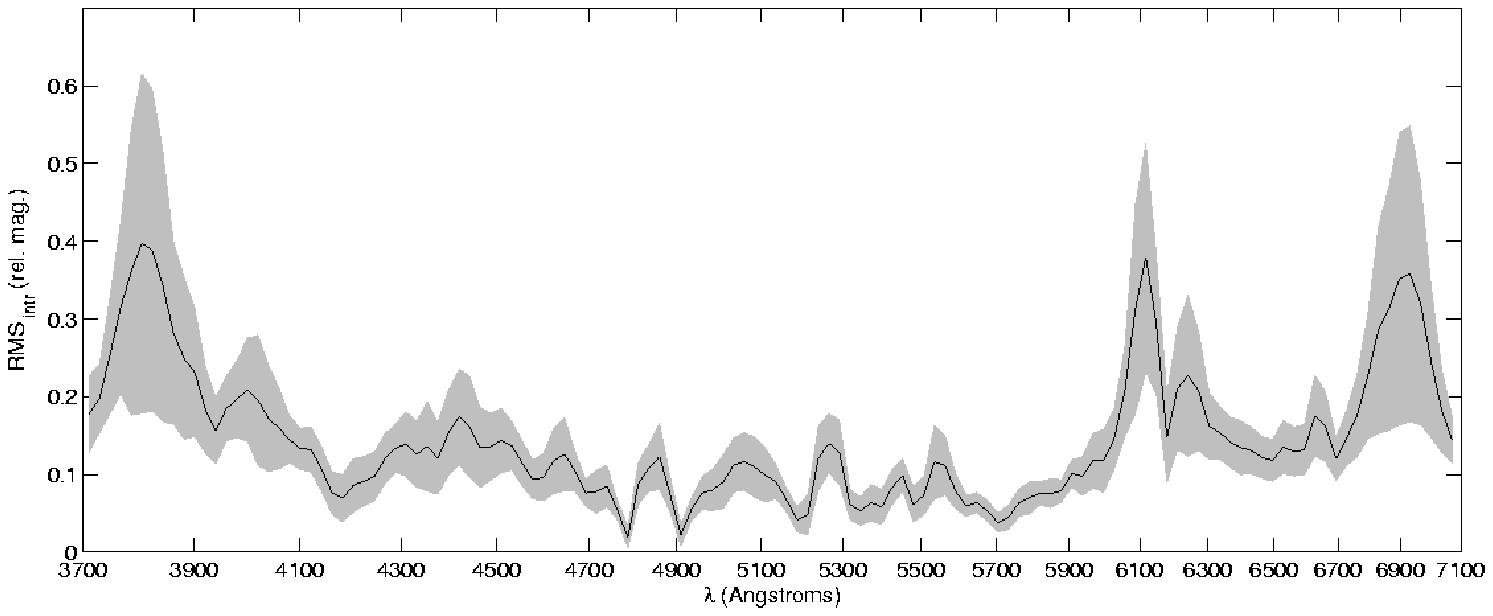}
\caption{Intrinsic RMS (total RMS minus observational dispersion) for spectra normalised to the average absolute magnitude in B (upper) and  V (lower).}\label{fig:fixedzoom}
\end{center}
\end{figure*}

 This method allows us to separate spectral diversity from scatter in absolute magnitudes.  It will be sensitive to diversity in spectral features as well as colour variaions --- large scale spectral shape differenes.  The band to which we scale the spectrum will have the least diversity because we are effectively anchoring the dispersion to be lowest at this point. So when we are scaling to a fixed $M_{\rm B}$ the scatter is smallest between 4000\AA\ and 5000\AA, whereas when scaling to $M_{\rm V}$ the scatter is smallest between 5000\AA\ and 6000\AA. The change in the overall shape of the dispersion spectrum is slight, and the significant spectral features remain unaltered. In Figure~\ref{fig:fixedzoom} we compare the final results for scaling to the V band (upper panel) and B band (lower panel).

\begin{figure}
\begin{center}
\includegraphics[]{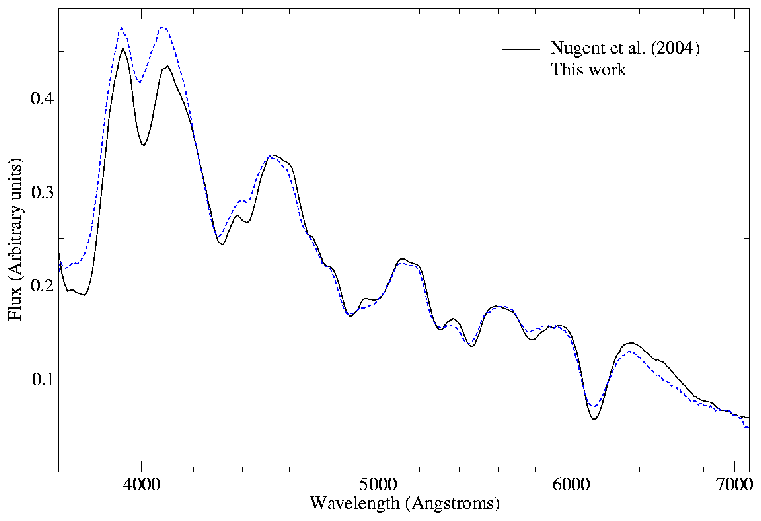}
\caption{The industry standard for typical SN Ia spectra are the templates provided in \citet{nugent02}.  These are updated periodically, and the update of April 2005 can be found on supernova.lbl.gov/$\sim$nugent/nugent\_templates.html.  Here we overplot our average spectrum on the Nugent template for a SN Ia at maximum light.  The agreement is good.  Our analysis shows that the UV edge of the spectrum, below the peak around 4100\AA, shows an increased diversity among SNe Ia.  (Whether this continues below the limit of our useful data, at 3700\AA, is an open question.)  This also approximately corresponds to where our average spectrum starts to deviate significantly from the Nugent template - corresponding to the greater uncertainty about the typical spectrum of this region.  We have shown that the Rayleigh-Jeans tail to the right of the peak is a very stable region of the spectrum, until we reach the Si II feature at 6150\AA.  This is therefore the best region of the spectrum to use as a standard candle for cosmological parameter estimation.}\label{fig:SNIamax}
\end{center}\end{figure}

In Figure~\ref{fig:SNIamax} we show a typical SN Ia spectrum at maximum light \citep{nugent02}. The spectrum peaks at 4000\AA. Our analysis shows that blueward of this peak the UV edge of the spectrum shows more variety than the rest of the spectrum.  Much of this is likely to be intrinsic SN Ia diversity, but it is also the region most susceptible to extinction and to observational flux errors.  Rejecting the blue end of SN Ia spectra will reduce the scatter in SN Ia magnitudes that occurs both due to intrinsic SN variability and imperfect extinction correction.  The Rayleigh-Jeans tail to the right of this peak, is a very stable region of the spectrum until the Si II feature around 6150\AA\ produces a peak in variability.   Beyond the Si II feature the diversity also appears to increase.  However, this may be another indication of imperfect extinction correction.  

We expect the dominant source of observational error that cannot be accounted for by measuring intra-object dispersion to be due to extinction correction.  By sampling the averaged object spectra with replacement many times we estimate the error in the extinction correction without making any assumptions about the error in colour excess values.   Because we anchor the spectra to the B or V magnitudes of the supernovae (even when we are anchoring to the B or V magnitude of each supernova individually), any faulty extinction correction would show up as a bowl-shaped curve on our RMS vs wavelength plots (affecting high and low wavelengths equally - even though extinction preferentially affects short wavelengths).
 The curves in Figure~\ref{fig:fixedzoom} may be the most illustrative for this point, since we have scaled the spectra to a fixed magnitude in B and V (upper and lower), effectively minimising the scatter in these regions.  An error in extinction correction shows up as a gradual increase in the scatter away from these regions.

In this analysis we are particularly concerned with (and sensitive to) small scale variation such as the variabiliy of a particular spectral feature.  Prominent spectral features correspond to regions of high scatter, most noticably around the Si II line at 6150\AA. Ideally we would avoid these regions when calculating the magnitudes of SNe Ia for use as standard candles.  However, a large variation in a spectral line does not necessarily translate into a large variation in broad-band magnitude if the regions around the line are bright and stable, in which case the variability of these lines will find the greatest utility as a tracer of different subsets of the SN Ia population.

In any real experiment we have a limited filter set, so cannot choose to observe the ideal region of every SN spectrum for the continuum of redshifts we need to cover.  However, we can weight the error bars of our observations to favour those objects whose redshifts fortuitously allow one of our filters to cover the most stable region, and we can design filters narrow enough, and with sufficient overlap, to maximise the chance that a filter will cover only the most stable region~\citep{davis05}.

Our analysis does not take into account any spectral differences between supernovae with different $\Delta m_{15}$ values.  When enough data is available it will be straightforward to apply this analysis to identify spectral features associated with wide or narrow light curves. 

\section[]{Measuring Peculiarity}\label{sect:peculiarity}
When using SNe Ia for cosmology we need to select the observed supernovae with the most homogeneous characteristics for use as standardisable candles.  Most well observed supernovae exhibit some peculiar features, and their full utility in cosmology requires the characterisation of these differences.  A significant step toward this is the creation of an objective measure of peculiarity that can act as an objective criterion for including an object in a sample population, and also identify subsets of SNe Ia with very similar properties. Here we outline two different methods for quantifying peculiarity and demonstrate them on the small data set currently available to us.  These methods will benefit greatly from a much larger low-$z$ spectroscopic sample, such as is now being collected by a number of groups, including the Nearby Supernova Factory, the Canada France Hawaii Telescope Legacy Survey (CFHTLS), the European Supernova Collaboration and the Carnegie Supernova Project \citep{hamuy05}.   In addition, applying these methods to the high-$z$ samples being collected by the ESSENCE collaboration, the CFHTLS and the SN Cosmology project will allow for tests of evolution in the demographics of the SN Ia population.   

The first method uses a $z$-test to determine the degree to which a SN Ia deviates from the norm, and is particularly good at finding spectra with variations in regions which are usually stable, or spectra with strange continuum profiles.  This means it is also good at picking out dubious extinction corrections.  The second method uses principle component analysis and is particularly good at finding subsets of the population with similar characteristics.  For example, it can pick out SNe Ia with high or low ejecta velocities by emphasising correlated differences between spectral features.

\subsection{Peculiarity Parameter}
We use our calculation of intrinsic diversity to give a measure of the range within which supernovae can be considered normal at each wavelength.  If an object deviates significantly from the average in a region of the spectrum which is very stable, such as the range from 4000\AA\ to 6000\AA, we might consider it `peculiar'. This can be quantified by assigning a $z$-score in each wavelength bin and then averaging the $z$-scores over some region, e.g. the entire spectrum or a band such as B or V. The $z$-score at wavelength $\lambda$ is,
\begin{equation}
z_{\lambda} = \frac{f - \bar{f}}{\sigma},
\end{equation}
where $f$, $\bar{f}$ and $\sigma$ are the flux of the object, the flux of the average spectrum and the dispersion of the average spectrum, all at the wavelength $\lambda$. We define the `peculiarity parameter', $P$, to be the absolute average of these terms over the $n$ bins in the desired wavelength range,
\begin{displaymath}
P = \frac{1}{n}\sum_{\lambda}|z_{\lambda}|.
\end{displaymath} 
Subsets of this range, weighted by filter value, can be used to define the broadband peculiarity parameters, e.g. for $B$,
\begin{displaymath}
P_B = \frac{1}{\sum_{\lambda}\tau_{\lambda}}\sum_{\lambda}\tau_{\lambda}|z_{\lambda}|,
\end{displaymath}
where $\tau_{\lambda}$ is the throughput of the $B$ band filter at wavelength $\lambda$.

\begin{figure}
\begin{center}
\includegraphics[]{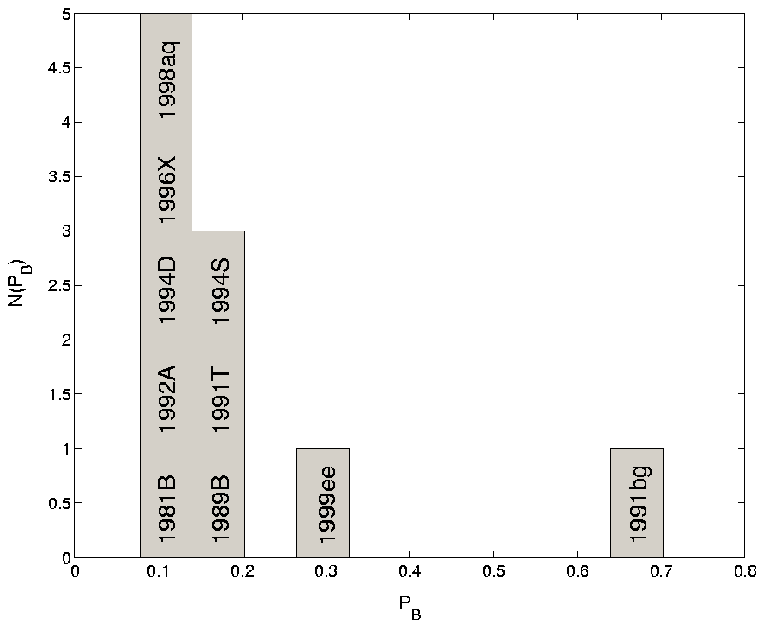}
\caption{Histogram of peculiarity parameter, $P$.  Spectra have been scaled to the average absolute magnitude in the B band.  The average spectrum from which deviation is measured is composed of average spectra from the objects listed excluding 1991bg and 1991T.}
\label{fig:pvalues_B}
\end{center}\end{figure}

Figure~\ref{fig:pvalues_B} shows a histogram of $P$ for the objects in our sample calculated according to this method using the wavelength range 3700 to 7100\AA, based on the object averages for which the spectra were anchored in the B band. For this test, the average and RMS spectra are recalculated including observations from SN~1991bg and SN~1991T, previously classified as `peculiar'. The average spectrum changes negligibly when these objects are added. The peculiarity of 1991bg is made obvious by the test, from which we conclude that the spectrum of this object departs markedly from the mean in a non-uniform (across $\lambda$) manner. However SN~1991T, which is known to be bolometrically overluminous, shows little spectral peculiarity when scaled in magnitude to match other events. 

\begin{figure}
\begin{center}
\includegraphics[]{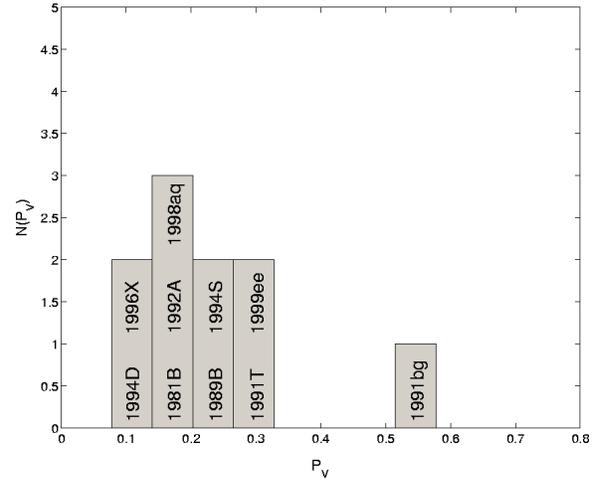}
\caption{Histogram of peculiarity parameter, $P$, with the average spectrum additionally excluding 1999ee.  Spectra have been scaled to the average absolute magnitude in the B band.  } 
\label{fig:pvalues_V}
\end{center}\end{figure}

To test the fidelity of this parameter, the histogram is compiled again using instead the spectra anchored in the V band. Figure~\ref{fig:pvalues_V} shows the recompiled histogram. Some slight rearranging occurs --- notably the objects seem less bunched --- however the prominence of 1991bg, and the relative lack of prominence of 1991T, are unchanged.  
We conclude that this parameter can distinguish between events that exhibit peculiar broadband and spectral features, such as 1991bg-like events, but does not distinguish objects like 1991T that are peculiar at wavelengths that also show significant diversity in normal supernovae.

\subsection{Principle Component Analysis}

Principle component analysis (PCA) is a technique for identifying common characteristics in data with many variables~\citep[e.g.][]{francis99}.  PCA has been successfully applied to astronomical spectra in the past to identify different types of active galactic nuclei~\citep{francis92}.  Here we demonstrate the concept on type Ia supernova spectra.  This technique may be used in the future, with a larger sample set, as another method to identify subclasses of type Ia supernovae with similar properties.  This is a major aim of several ongoing low-redshift supernova surveys and is important for the control of systematic errors in planned (and ongoing) high-redshift surveys that aim to measure the expansion history of our universe.
We use the PCA technique developed by~\cite{francis92} including their analysis code with some minor variations.\footnote{Many thanks to Paul Francis for supplying us with this code and running the first analysis.}  

\begin{figure}
\begin{center}
\includegraphics[]{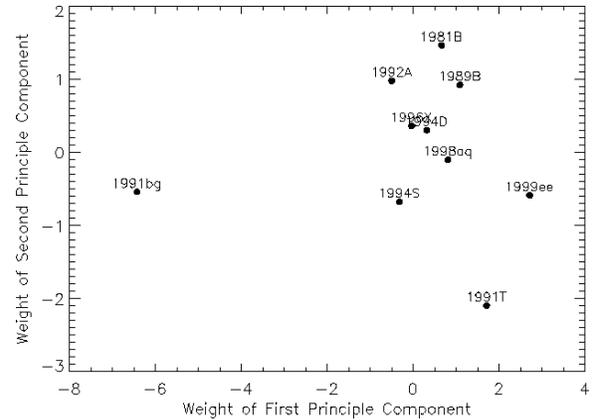}
\caption{For each SN spectrum we plot the weights of the first and second components of a principle component analysis.  The underluminous event, SN1991bg, is identified as peculiar in the first principle component, and the overluminous SN 1991T is the most peculiar SN in the second principle component.  This is particularly interesting because the spectra have been scaled to the same overall flux between 3700\AA\ and 7100\AA, so the peculiarity that is apparent here is a difference in the shape and features of the spectrum, rather than simply a shift in the luminosity.}
\label{fig:PCAallsame}
\end{center}\end{figure}

We perform PCA on the corrected sample of normal supernovae as outlined in Table~\ref{table:spectra}, as well as extinction-corrected spectra for the peculiar SNe 1991T and 1991bg.  The result is shown in Fig.~\ref{fig:PCAallsame}, where we have plotted co\"efficients of the first and second principle components, calculated for each supernova. This test successfully identifies these two supernovae as peculiar.  Moreover, this method is able to distinguish 1991bg-like SNe from 1991T-like SNe because they each appear peculiar in a different principle component.  This method therefore shows great promise for selecting sub-classes of SNe Ia.  Future work will perform this analysis on multi-epoch spectra, in which sub-class trends should be even more prominent.

\section{Conclusions}
We have used published spectral data to calculate the diversity of type Ia supernova spectra close to maximum light as a function of wavelength between 3700\AA\ and 7100\AA.  While results in the regions toward the UV and NIR are sensitive to systematics of our calibration methodology, the diversity is particularly stable between 4100\AA\ and 6000\AA\ in the rest frame of the supernova.  This is therefore the ideal region for calculating SN Ia magnitudes for use as standard candles. There is an increase in supernova diversity below 4100\AA.  As we move blueward, extinction correction also becomes more uncertain and therefore the region below rest frame 4100\AA\ should be avoided when using SN Ia as standardizable candles.  Observations of this region remain useful for constraining theoretical models of SN Ia explosions, for example \cite{nugent95} showed the U band is a powerful tool for discriminating between different processes occurring in SNe Ia. Redward of the stable region the Si II line, typically observed at 6150\AA, shows a large diversity.  This feature has previously been used as a tracer of light-curve width \citep{nugent95} and to identify peculiar SNe Ia~\citep{bra93a,bra93b}.  Our analysis confirms that it is a variable feature of ordinary SNe Ia.  Future work will extend this study to spectra over a wider epoch range. 

We also demonstrated two methods for determining the peculiarity of type Ia supernovae, firstly by comparing their spectra to the average SN Ia spectrum while taking into account the typical variability of different spectral regions, and secondly using principle component analysis. Both methods show promise for use as objective tests of peculiarity.  PCA in particular may be a good way to objectively define sub-types of the SN Ia class. 

The methods demonstrated here will benefit greatly from more spectroscopic data, not only more objects, but also extended wavelength coverage into the UV and IR.  Although hundreds of supernovae have now been observed very well photometrically, the sample that have spectra published near maximum remains small.  New data soon to be available from a variety of searches will enable a more robust analysis based on the methods demonstrated here. Such an analysis would aim to understand the nature, rather than the extent, of the diversity in the population. Potential sources include variations in expansion velocities, chemical composition and \emph{ex-situ} factors such as evolution and gravitational lensing. Relationships between separate supernova events could then be drawn by a study of correlations in spectral residuals. This subject will be tackled in future work.

\section*{Acknowledgments}
Many thanks are due to Paul Francis for sharing with us his principle component analysis code.  We also thank the many observers who are responsible for obtaining this data.  JBJ was supported by an Australian National University Summer Research Scholarship.  TMD acknowledges the support of Lawrence Berkeley Laboratory. We thank the anonymous reviewer for their helpful comments.

\bsp
\label{lastpage}

\begin{thebibliography}{99}
\bibitem[Benetti et al.(2004)]{benetti04}Benetti, S.~et al.~2004, MNRAS, 348, 261
\bibitem[Benetti et al.(2005)]{ben05}Benetti, S.~et al.\ 2005, ApJ, 623, 1011
\bibitem[Bongard et al.(2006)]{bongard06}Bongard, S.\ et al.\ 2006, astro-ph/0512229
\bibitem[Blondin et al.(2004)]{blondin04}Blondin, S.\ et al.\ 2006, AJ, in press, astro-ph/0510089
\bibitem[Branch et al.(1983)]{bra83}Branch, D.~et al.~1983, ApJ, 270, 123 (Br83) 
\bibitem[Branch et al.(1993a)]{bra93a}Branch, D., van der Bergh, S.~1993, AJ, 105, 2231
\bibitem[Branch et al.(1993b)]{bra93b}Branch, D., Fisher, A., \& Nugent, P.~1993, AJ, 106, 2383
\bibitem[Branch et al.(2003)]{bra03}Branch, D.~et al.~2003, AJ, 126, 1489 (Br03)
\bibitem[Branch et al.(2005)]{branch05}Branch, D., Baron, E., Hall, N., Melakayil, M.\ and Parrent, J.\ 2005, PASP, 117, 545
\bibitem[Branch et al.(2006)]{branch06}Branch, D.\ et al.\ 2006, astro-ph/0601048
\bibitem[Davis et al.(2006)]{davis05}Davis, T.~M., Schmidt, B.~P., \& Kim, A.~2006, PASP, {\em in press}
\bibitem[Efron (1982)]{efron82}Efron, B., 1982, \emph{The Jackknife, the Bootstrap and other resampling plans}, CBMS-NSF Regional Conference Series in Applied Mathematics, Philadelphia: Society for Industrial and Applied Mathematics (SIAM)
\bibitem[Filippenko(1997)]{filippenko97}Filippenko, A.~V.\ 1997, ARAA, 35, 309
\bibitem[Francis et al.(1992)]{francis92}Francis, P.~J., Hewett, P.C., Foltz, C.B. \& Chaffee, F.H.\ 1992, ApJ, 398, 476
\bibitem[Francis and Wills(1999)]{francis99}Francis, P.~J.\ \& Wills, B.~J.\ 1999, Quasars and Cosmology, ASP Conference Series 162,  Eds Ferland, G. \& Baldwin, J., p.~363
\bibitem[Gallagher et al.(2005)]{gallagher05}Gallagher, J.~S.\ et al.\ 2005, ApJ, 634, 210
\bibitem[Hamuy et al.(1996)]{hamuy96}Hamuy, M., Phillips, M.~M., Schommer, R.~A., Suntzeff, N.~B., Maza, J., \& Aviles, R.\ 1996, AJ, 112, 2391
\bibitem[Hamuy et al.(2002)]{hamuy02}Hamuy, M.~et al.~2002, AJ, 124, 417
\bibitem[Hamuy et al.(2005)]{hamuy05}Hamuy, M.~et al.~2005, PASP, in press
\bibitem[Jha et al.(2005)]{jha05}Jha, S.\ et al.\ 2005, AJ, 131, 527
\bibitem[Kirshner et al.~(1993)]{kir93} Kirshner, R.~P., et al.\ 1993, ApJ, 415, 589 (Ki93)
\bibitem[Knop et al.(2003)]{knop03}Knop, R.~A. et al.\ 2003, ApJ, 598, 102 (Kn03)
\bibitem[Leibundgut et al.(1993)]{lei93}Leibundgut, B.~et al.~1993, AJ, 105, 301 (Le93)
\bibitem[Nugent et al.(1995)]{nugent95}Nugent, P., Phillips, M., Baron, E., Branch, D. and Hauschildt, P.~1995, ApJL, 455, L147
\bibitem[Nugent, Kim \& Perlmutter(2002)]{nugent02}Nugent, P., Kim, A., \& Perlmutter, S.~2002, PASP, 114, 803
\bibitem[Patat et al.(1996)]{pat96}Patat, F., Benetti, S., Cappellaro, E., Danziger, I.~J., della Valle, M., Mazzali, P.~A., \& Turatto, M.\ 1996, MNRAS, 278, 111 (Pa96)
\bibitem[Phillips(1993)]{phil93}Phillips, M.~M.\ 1993, ApJL, 413, L105 
\bibitem[Phillips et al.(1999)]{phil99}Phillips, M.~M., Lira, P., Suntzeff, N.~B., Schommer, R.~A., Hamuy, M., \& Maza, J.\ 1999, AJ, 118, 1766 (Ph99)
\bibitem[Saha et al.(2001)]{sah01}Saha, A., Sandage, A., Tammann, G.~A., Dolphin, A.~E., Christensen, J., Panagia, N., \& Macchetto, F.~D.~2001, ApJ, 562, 314 (Sah01)
\bibitem[Salvo et al.(2001)]{sal01}Salvo, M.~E., Cappellaro, E., Mazzali, P.~A., Benetti, S., Danziger, I.~J., Patat, F., \& Turatto, M.\ 2001, MNRAS, 321, 254 (Sal01)
\bibitem[Savage \& Mathis(1979)]{savage79}Savage, B.~D., \& Mathis, J.~S.\ 1979, ARA\&A, 17, 73
\bibitem[Stritzinger et al.(2002)]{stri02}Stritzinger, M.~et al.\ 2002, AJ, 124, 2100 (St02)
\bibitem[Travaglio et al.(2005)]{travaglio05}Travaglio, C., Hillebrandt, W.\ and Reinecke, M.\ 2006, AAP, in press, astro-ph/0507510
\bibitem[Wells et al.(1994)]{wel94} Wells, L.~A.~et al.\ 1994, AJ, 108, 2233 (We94)
\end{thebibliography}
\end{document}